# Two-dimensional LWR model for lane-free traffic

**Shrey Agrawal**
Department of Civil Engineering
Indian Institute of Technology Kanpur
Email: shrey20@iitk.ac.in

**Venkatesan Kanagaraj***
Department of Civil Engineering
Indian Institute of Technology Kanpur
Email: venkatk@iitk.ac.in

and

**Martin Treiber**
Institute for Transport and Economics
Technical University Dresden, Germany
Email: martin.treiber@tu-dresden.de

**\* Corresponding Author**

**Abstract**

While macroscopic models for single or multi-lane traffic flow are well established, these models are not applicable to the dynamics and characteristics of disordered traffic which is characterized by widely different types of vehicles and no lane discipline. We propose a first-order two-dimensional Lighthill-Whitham-Richards (LWR) model for the continuous macroscopic longitudinal and lateral dynamics of this type of traffic flow. The continuity equation is extended into two dimensions and the equation is closed by assuming a longitudinal flow-density relationship as in traditional one-dimensional models while the lateral dynamics is based on boundary repulsion and a desire of a majority of the drivers to go to less dense regions. This is equivalent to Fick's law giving rise to a lateral diffusion term. Using the proposed model, several numerical tests were conducted under different traffic scenarios representing a wide range of traffic conditions. Even for extreme initial conditions, the model's outcome turned out to be plausible and consistent with observed traffic flow dynamics. Moreover, the numerical convergence test is performed using an analytical solution for lateral steady-state conditions. The model was applied for bicycle simulation and reproduced the evolution of lateral density profile with asymmetric behavior.

# 1. Introduction

Macroscopic traffic flow models represent the traffic state in terms of aggregate quantities like flux, density, and average local speed, and describe how these variables evolve over space and time. For ordered lane-based traffic, numerous continuum traffic flow models are existing and well-tested. The first macroscopic traffic flow models were proposed by Lighthill and Whitham (1955) and Richards (1956) independently using kinematic wave theory. The resulting LWR model class consists of three main parts: a hydrodynamic relationship, a conservation equation, and an equilibrium speed-density relationship as shown below:

$$\frac{\partial \rho}{\partial t} + \frac{\partial Q}{\partial x} = 0 \qquad (1)$$

with $Q = \rho v$ and $v = v(\rho)$

Furthermore, some researchers introduced lane changing process in macroscopic models. One of the earliest works from this perspective is by Gazis et al. (1962) which was further improved by Munjal and Pipes (1971) who assumed lane changing will occur to attain uniform density across the road section. Michalopoulos et al. (1984) proposed a second-order model with lane changing as a function of momentum difference. Daganzo (2002) came up with a behavioral theory for multi-lane traffic by defining two possibilities (i) two pipe regime for slugs and rabbits with a passing lane having higher velocity than the shoulder lane, and (ii) a single pipe regime for all the vehicles traveling at speed lower or equal to slug free-flow speed. Laval and Daganzo (2006) quantified the lane-changing rate and then applied it to the hybrid model considering lane changers as moving bottlenecks with bounded acceleration. Roncoli et al. (2015) proposed a lane-changing model using density difference as attractiveness for the adjacent lane, and further, added a weighting to reflect particular location and time-dependent effects wherever needed. Nagalur Subraveti et al. (2019) improved the Roncoli et al. (2015) model by introducing incentive-based weightage, where incentives are density difference, keep-right bias, route, and courtesy. This allows the model to take effects such as lane drop, off-ramp, on-ramp, and compulsory lane changing. Most of the first-order multi-lane models assume road width is divided into discrete lanes and lateral flows are calculated at a discrete level. In principle, a spatially two-dimensional formulation should better describe the traffic flow process.

In disordered traffic conditions, limited attempts were made to consider different vehicle types and dynamics in the longitudinal direction. Nair et al. (2011) proposed a continuum model for disordered and heterogeneous traffic stream considering such a traffic stream as a porous medium and considering speed–density relationship based on pore space distribution. Shiomi et al. (2015) proposed a two-class two-phase (creeping and non-creeping) model considering the creeping behavior of smaller-sized vehicles during congested periods when larger vehicles come to a complete stop. Gashaw et al. (2018) developed a model that describes the impacts of cars and powered two-wheelers. The flow of two-wheelers is considered as a fluid in a porous medium and vehicle spacing is considered as a distribution. The probability density function for vehicles' spacing distribution is estimated from multiple simulation runs of traffic flow with different percentage compositions of vehicles. Bhavathrathan and Mallikarjuna (2012) developed a macroscopic traffic flow model incorporating the side-by-side movement of two-wheelers using the concept of diminishing density. Mayakuntla and Verma (2019) proposed a cell transmission model for heterogeneous traffic. In the model, two types of classes are considered such as car-following and gap-filling from which two forms of fundamental relationships are derived.

Ahmed et al. (2019) proposed a width-based cell transmission model developing the flow-density relationship using parameters like road width. Gupta and Dhiman (2014) developed a higher-order model for non-lane-based traffic using the full-velocity difference car-following model by taking into account the lateral separation of vehicles. Mohan and Ramadurai (2021) extended the second-order AR model proposed by Aw and Rascle (2000) for multi-class traffic on a three-dimensional flow concentration surface. In this model, the flow of vehicle class is a function of class density and the fraction of road area occupied by other vehicle classes. Fosu et.al. (2021) developed a new viscous second-order macroscopic model. In this model, they introduced the viscosity term and velocity gradient in the lateral direction in the longitudinal momentum equation. Kaur et al. (2022) developed a lattice hydrodynamic area occupancy model for disordered traffic. In this model, the road section is longitudinally divided into lattices, and traffic dynamics are updated based on the conservation of vehicles and flow evolution equations. Different types of vehicles are introduced in the model based on the area occupancy concept. The above-mentioned studies cover the research topics such as macroscopic models and multi-lane models for disordered traffic. All the above-mentioned models deal with longitudinal dynamics of the traffic. But to represent disordered traffic more precisely, there is a need to extend the continuum of traffic dynamics in two dimensions to capture lateral movements.

There are some first propositions to develop 2D continuum models. Herty et al. (2018) proposed a novel two-dimensional first-order macroscopic model using experimental traffic data obtained from Germany and US. In this model, longitudinal and lateral traffic variables are continuous. The equilibrium speed-density relationship was obtained using a regression line on fundamental diagrams. In this study, lateral dynamics is rather trivial and not realistic, for example, it is always negative (pointing towards the right), very small, and does not depend on the gradients to guide traffic flow towards regions of lower density. In the simulations, the lateral dynamics shape thus remains essentially unchanged. Moreover, the simulation time is less than 1 s. Balzotti and Goettlich (2021) extended the two-dimensional multi-class model from the single-class model. (Herty et al. 2018) analyzed the two-dimensional Riemann problem and tested it with real traffic data. In this paper, the authors transformed three micro-vehicles (two cars, and one truck) into macroscopic quantities by kernel smoothing. However, for just three microscopic objects, a macroscopic approach is not justifiable. Moreover, the macroscopic truck envelope just vanishes from the road which is again a sign that no proper boundary conditions are used. Herty et al. (2018) proposed a two-dimensional macroscopic second-order model which accounts for lane-changing behaviour. In this study, the lane is considered as a continuum medium and captures traffic dynamics caused by lane-changing manoeuvers. However, again, the lateral flow is minuscule (therefore it is scaled in multiples of $10^4$ in the figures) and the simulation time (3 seconds) is minimal without any visible lateral dynamics. Hence, lateral dynamics is rather trivial and not realistic.

Sukhinova et al. (2009) proposed a hybrid first-order (lateral - y direction) and second-order (longitudinal - x direction) macroscopic model for multi-lane traffic considering the following factors: accelerating/decelerating force, density gradient, the driver trying to change faster lane, vehicle's speed. However, this model is significantly more complicated and has no boundary repulsion forces. Mohan (2019) proposed a multi-class two-dimensional AR model for lane-less traffic. Vikram et al. (2022) proposed two-dimensional higher-order models for disordered traffic. In the higher-order model proposed by this study, longitudinal acceleration is considered as the function of density and density gradient in the longitudinal direction, and lateral acceleration is considered as the function of density gradient in the lateral direction. The above-mentioned studies do not consider the proper boundary-repulsive forces. Mollier et al. (2019) proposed a two-dimensional macroscopic model on a network level. In this model, velocity magnitude is given by a macroscopic fundamental diagram (MFD) while the velocity direction is constructed using the network geometry. Notice that this MFD model does not consider the lateral dynamics of vehicles on the road level.

In summary, from the literature review, we conclude that the research on two-dimensional first and second-order macroscopic traffic flow models has just started but none of the investigations mentioned above came up with a fully two-dimensional model including the effect of road boundaries and lateral density gradients. Regarding the model order, we observe that a fully two-dimensional second-order model is mathematically complex. For example, it consists of three equations for the density and the two velocity components and both equations for the velocity components include separate advective, relaxation, and interaction terms. This leads to many parameters, a demanding numerical simulation, and makes such prospective models hard to calibrate. While first-order (LWR) models cannot describe traffic instabilities, they are still one of the most powerful models in describing traffic breakdown and congestion behind bottlenecks such as on-ramps, lane drops, traffic lights, and incidents. Furthermore, they can be easily simulated and extended to network traffic using the supply-demand method. Moreover, first-order models are also analytically tractable and have fewer parameters, and are easier to calibrate than second-order models. To our knowledge, a complete first-order 2D macroscopic model for lane-free traffic including boundary repulsion, avoiding obstacles, managing bottlenecks, and modeling the desire of drivers to go to less dense regions is missing.

In this contribution, we propose a complete first-order 2D macroscopic model for directed lane-free traffic by generalizing the conventional continuity equation to two dimensions and including lateral dynamics driven by diffusion and boundary repulsion. The diffusion term is motivated by Fick's law: Even without a driver's desire, the unordered lateral motion will lead to a net flow away from density clusters, and this lateral flow is enhanced by the driver's desire. In lane-based traffic, this corresponds to the human desire to use emptier lanes. The boundary repulsion reflects the fact that, unlike the longitudinal dimension, the lateral dimension is physically restricted by the road width and reflects the desire of the drivers to keep on the road surface. Simulations show that the model reproduces the observed characteristics of disordered traffic. In the next two sections, we specify the model and propose a numerical update scheme. In section 4, we simulate the model with a wide range of initial conditions representing different, partially extreme, traffic scenarios before we conclude in section 5.

## 2. Model Specification

In disordered traffic without lanes or lane discipline, the longitudinal dynamics are similar to lane-based traffic. Simultaneously, the drivers move laterally to occupy free space wherever it is available. Hence, in the first-order macroscopic model, both longitudinal and lateral dynamics are considered together. In this study, we proposed a 2D LWR model by extending the 1D continuity equation into two- dimensions as follows:

$$\frac{\partial \rho}{\partial t} + \frac{\partial Q_x}{\partial x} + \frac{\partial Q_y}{\partial y} = 0, \forall t > 0, x \in [0, L], y \in [-b, b] \quad (2)$$

The density $\rho(x, y, t)$ is defined as the number of vehicles per square area [veh/m²], and $Q_x$ and $Q_y$ represent flux density [veh/ms] in the longitudinal (x) and lateral (y) directions, respectively. Unlike pedestrian traffic, 2D traffic flow is clearly anisotropic with longitudinal dynamics distinctly different from the lateral ones. Hence, different expressions are used for the longitudinal and lateral flow-density in our model. The flux density function based on hydrodynamic relationship can be expressed as:

$$Q_x = \rho v_x \quad (3)$$

$$Q_y = \rho v_y - D \frac{\partial \rho}{\partial y} \quad (4)$$

where $v_x$ and $v_y$ are the x and y coordinates of the velocity, respectively and $D$ is the diffusion coefficient. Moreover, the flux density function in the lateral direction (equation 4) is proportional to an additional term, density gradient. This is motivated by Fick's law: Random lateral motion and the desire of drivers leads to a net lateral flow along the negative gradient of density. For the first-order model, this means lateral flow density, $Q_y$ is proportional to $D \frac{\partial \rho}{\partial y}$. Including this into the continuity equation gives a diffusion term in equation (8).

**Longitudinal Dynamics:** First, a relationship needs to be established between density and speed in the x-direction. In the 1D LWR model, speed is a decreasing function of the density with the free-flow speed at zero density and zero speed at the maximum (jam) density. Different functional relations have been proposed for the speed-density relationship. In this paper, we used Greenshields' model and Del Castillo and Benitez's model (Castillo and Benítez, 1995) for longitudinal movement. Greenshields' model, which arguably is the simplest model satisfying the conditions above:

$$v_x = v_f \left[ 1 - \frac{\rho}{\rho_{max}} \right] \quad (5)$$

Here, $v_f$ is the free-flow speed in the x-direction and $\rho_{max}$ is the jam density. Notice that, unlike Greenshields' model, we apply it for two-dimensional densities [veh/m²]. Del Castillo and Benitez (1994) proposed the following speed-density exponential functional form:

$$v_x = v_f \left[1 - exp\left[1 - exp\left(\frac{|C_j|}{v_f}\left(\frac{\rho_{max}}{\rho} - 1\right)\right)\right]\right] \qquad (6)$$

Where $|C_j|$ = kinematic wave speed at jam density. Free-flow speed and jam density are estimated using empirical data collected in disordered traffic conditions (Ahmed et al. 2021). Empirical data shows that the propagation velocity of approximately $|C_j|$ = 15 km/h is universal in congested traffic situations (Treiber and Kesting, 2013). Table 1 shows the parameter values of two longitudinal models. The speed vs density, flow–density vs density and characteristics speed vs density for the given model parameters are shown in Figure 1. These parameter values of both models are used for further analysis.

Table 1: Longitudinal Model Parameters

| Parameters | Greenshields' Model | Del Castillo and Benitez's model |
|---|---|---|
| Jam Density, $k_j$ | 0.062 veh/m² ||
| Free flow longitudinal speed, $v_f$ | 15.28 m/s ||
| Magnitude of kinematic wave speed at jam density, $|C_j|$ | Not Applicable | 4.16 m/s |

**Lateral Dynamics:** Unlike the longitudinal dimension, the lateral dimension is physically restricted by the road width. In order to capture the resulting repulsive boundary effects, we introduced a lateral desired speed which exponentially increases when approaching either boundary of the road:

$$v_y = v_{yb}\left[e^{\left(-\left(\frac{b+y}{s_y}\right)\right)} - e^{\left(-\left(\frac{b-y}{s_y}\right)\right)}\right]\left(1 - \frac{\rho}{\rho_{max}}\right) \qquad (7)$$

Where +/- $v_{yb}$ denotes the desired lateral speed immediately at one of the boundaries, b is the half-width of the road, $s_y$ is the ''road boundary scale'' which is a model parameter, and y is the lateral distance from the road center. In equation (7), a maneuverability factor $\left(1 - \frac{\rho}{\rho_{max}}\right)$ act as a directed lateral movements (equivalent of lane changing criteria for discrete lanes). If this term is not present in the equation, vehicles start with a constant maximum density as initial condition and later, the boundary repulsion forces drive the density above the maximum inside the road. So, this term serves for modelling consistency, i.e., if the maximum density is reached, no vehicle can move away from the boundary as there is no space.

A second macroscopic lateral flow is caused by density gradients. From a gas-kinetic point of view, each vehicle-driver unit has a random lateral velocity component which, even without an explicit driver's desire, leads to a net lateral flow away from the lateral density gradient (Fick's law) and to a lateral diffusion term. This effect is augmented by the explicit desire of some drivers to go to less dense regions. With the lateral repulsion and diffusion terms, the final 2D LWR model equations read

$$\frac{\partial \rho}{\partial t} + \frac{\partial (\rho v_x)}{\partial x} + \frac{\partial (\rho v_y)}{\partial y} = D \frac{\partial^2 \rho}{\partial y^2}, \forall t > 0, x \in [0, L], y \in [-b, b] \quad (8)$$

with (5) - (7). Notice that the lateral diffusion also helps avoid sharp transitions and shocks in the lateral direction.

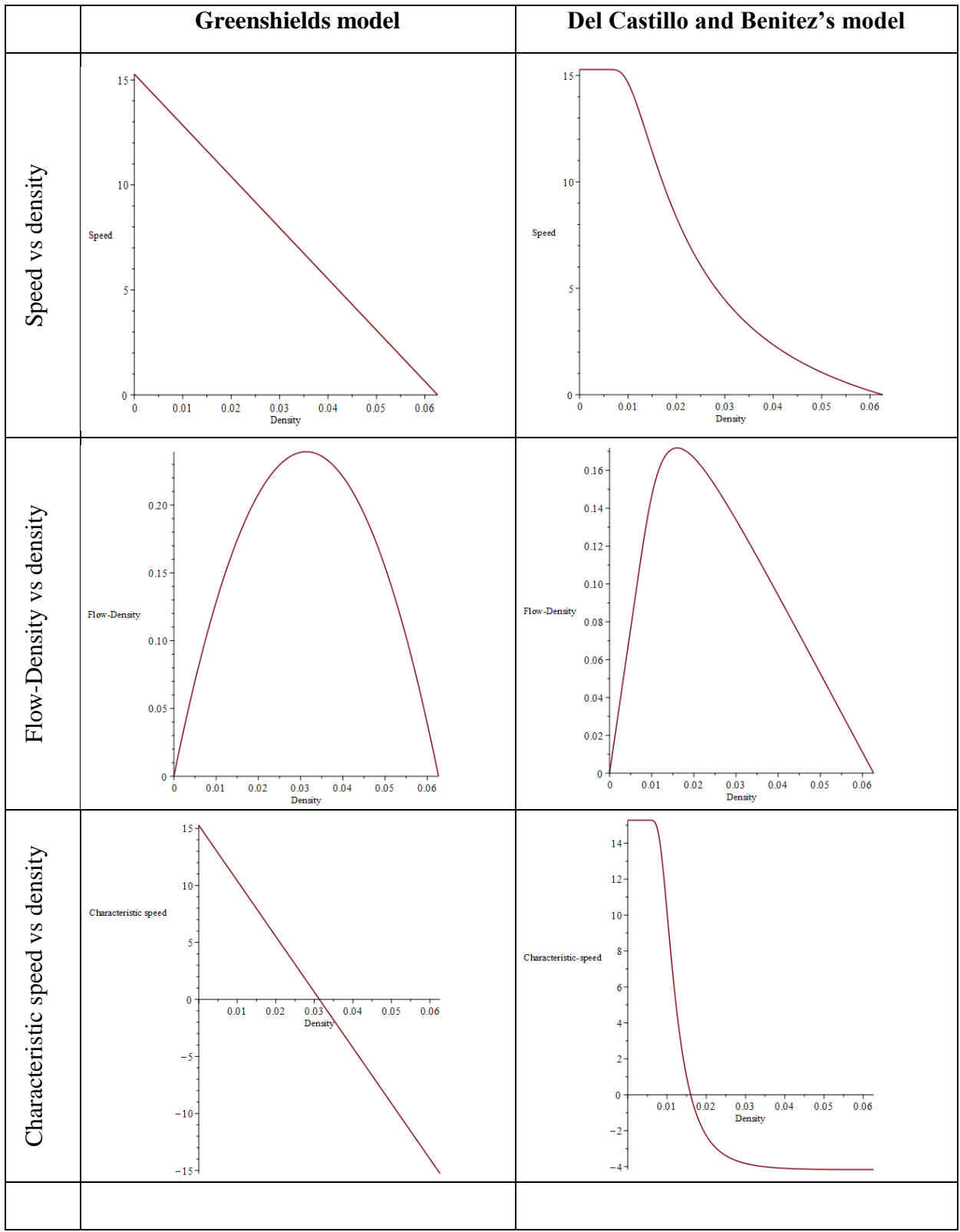

Figure 1: Speed vs density, flow–density vs density and characteristics speed vs density Diagrams

## 3. Numerical Scheme

In this section, a numerical scheme to solve the proposed 2D LWR model is discussed. The conservation equation with diffusion term possible can be modelled using numerical methods based on directional splitting (Strang 1968). The equation of the model is split into three different parts and are computed separately:

$$\frac{\partial \rho}{\partial t} + \underbrace{\frac{\partial Q_x}{\partial x}}_{I\ term} + \underbrace{\frac{\partial \widetilde{Q_y}}{\partial y}}_{II\ term} = \underbrace{D \frac{\partial^2 \rho}{\partial y^2}}_{III\ term} \quad (9)$$

Where $\widetilde{Q_y} = \rho v_y$. In the first step, propagation of density in the x coordinate is computed and the propagation of density in the y coordinate is updated in the second step. Finally, the diffusion term is accounted using operator splitting method (Toro, 2013; Gosse, 2014). Spatial domain discretization is represented graphically in Figure 2 which depicts the flow interface in the x and y axis. In the longitudinal direction, it is assumed that the velocity ranges from zero to free flow speed and is always in the forward direction. Therefore, the respective longitudinal flux is unidirectional ranging from zero to capacity flow values. While, in the lateral direction, flux is bidirectional which moves either left to right or right to left direction based on whether the lateral velocity is negative or positive. Flux is computed using the Gudonov scheme for both x and y directions and the diffusion term is computed using the central difference method. In numerical context, we want to evolve the solution from its initial value $\rho^n$ at time $t^n$ to the next time step i.e., $\rho^{n+1}$ at time $t^{n+1}$, given by $t^{n+1} = t^n + \Delta t$. For that, spatial domain is discretised into finite cells (i, j) in x and y-direction.

The numerical flux is defined as the interface between two cells $Q_{i+\frac{1}{2},j} = Q(\rho_{i,j}, \rho_{i+1,j})$ and longitudinal flux is calculated using the following equation:

$$Q_{i+\frac{1}{2},j} = \Phi_x\left(\rho_{i+\frac{1}{2},j}\right) = \begin{cases} \min(\Phi_x(\rho)); & \rho_{i,j} < \rho_{i+1,j} \\ \max(\Phi_x(\rho)); & \rho_{i,j} \geq \rho_{i+1,j} \end{cases} \quad (10)$$

Here $\Phi_x$ represent the flux function in the longitudinal direction and the flux direction is downstream in the longitudinal direction. The lateral flux is defined as the following equation:

$$Q_{i,j+\frac{1}{2}} = \Phi_y\left(\rho_{i,j+\frac{1}{2}}\right) = \begin{cases} \min\left(\Phi_y(\rho)\right); & \rho_{i,j} < \rho_{i,j+1} \\ \max\left(\Phi_y(\rho)\right); & \rho_{i,j} \geq \rho_{i,j+1} \end{cases} \quad (11)$$

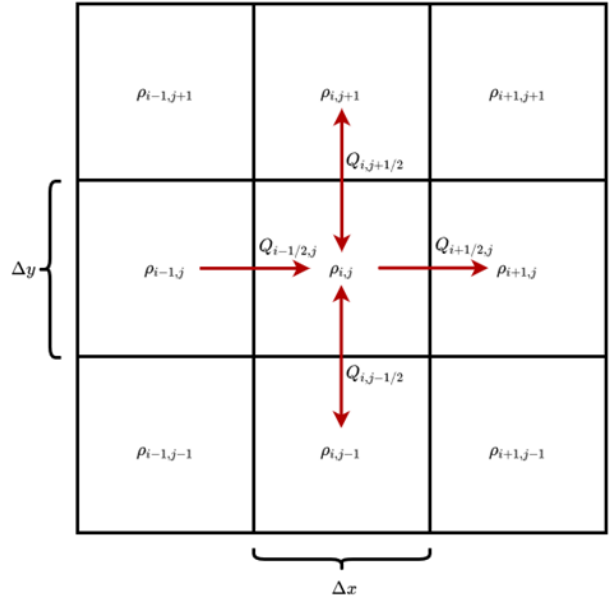

Figure 2: Spatial domain discretization into finite cells

Here $\Phi_y$ represents the flux function in the lateral direction and the flux direction is based on density in the adjacent cells. The overall scheme for the computation of the model is given below:

$$\rho_{i,j}^* = \rho_{i,j}^n + \left(Q_{i+1/2,j}^n - Q_{i-1/2,j}^n\right)\frac{\Delta t}{\Delta x} \tag{12}$$

$$\rho_{i,j}^{**} = \rho_{i,j}^* + \left(Q_{i,j+1/2}^* - Q_{i,j-1/2}^*\right)\frac{\Delta t}{\Delta y} \tag{13}$$

$$\rho_{i,j}^{n+1} = \rho_{i,j}^{**} + D\left(\rho_{i,j+1}^{**} - 2\rho_{i,j}^{**} + \rho_{i,j-1}^{**}\right)\frac{\Delta t}{\Delta y^2} \tag{14}$$

The time step is $\Delta t$, discretization of space with respect to x-axis and y-axis are $\Delta x$ and $\Delta y$, receptively. The Courant–Friedrichs–Lewy (CFL) conditions pose a time step restriction to guarantee the stability of the numerical scheme. The first CFL criteria relate to convective instability caused by first-order spatial derivatives and read

$$\Delta t \leq \frac{\Delta x}{v_f} \quad \& \quad \Delta t \leq \frac{\Delta y}{v_{yb}} \tag{15}$$

and the second CFL criterion to avoid diffusive instability is given by

$$\Delta t \leq \frac{(\Delta y)^2}{2D} \tag{16}$$

Finally, the time step adopted is less than or equal to time step which is given by CFL criteria as follows:

$$\Delta t \leq min\{\Delta x/v_f, \Delta y/v_{yb}, (\Delta y)^2/2D\} \tag{17}$$

## 4. Numerical Test

In order to evaluate the proposed model discussed in section 2, numerical tests were conducted under different traffic scenarios which represent a wide range of traffic conditions such as concave and convex densities in the lateral direction, shock waves, and rarefaction waves. All the numerical tests were conducted for a uniform road section of length $L$ and width $2b$ defined on $[0, L] \times [-b, b]$. There is no lateral access and egress of vehicles, so the lateral flux is zero at the lateral road boundaries. At the longitudinal upstream and downstream boundaries, we have assumed Von-Neuman boundary condition $\frac{\partial \rho}{\partial x} = 0$ reflecting an unperturbed inflow and outflow. Longitudinal model parameters are used from the Table 1 in section 2. The assumed parameters are given in Table 2. Notice that the first two are infrastructure parameters, the next four are numerical parameters, and the rest are actual model parameters.

Table 2: Parameters considered in this study

| Parameter | Value |
|---|---|
| Road length, L | $1000\ m$ |
| Road width, $2|b|$ | $10\ m$ |
| Longitudinal step-size, $\Delta x$ | $10\ m$ |
| Lateral step-size, $\Delta y$ | $0.1\ m$ |
| Mesh $N^x \times N^y$ | $100 \times 100\ cells$ |
| Time step-size, $\Delta t$ | $0.001\ s$ |
| Lateral boundary interaction, $s_y$ | $2\ m$ |
| Maximum boundary repulsion velocity, $v_{yb}$ | $0.5\ m/s$ |
| Coefficient of diffusion, $D$ | $1\ m^2/s$ |

**Steady-state lateral density profile:**

In this study, Gudonov scheme is used for convection term and central difference method is used for diffusion term in lateral direction based on directional splitting. The implementation and correctness of numerical scheme of lateral convection and diffusion terms is checked by comparing with analytical solution of lateral-steady state distributions. In this regard, first, we have derived the analytical solution for steady-state density distribution. For that, we assumed steady state $\frac{\partial}{\partial t} = 0$ and homogenous roads $\frac{\partial}{\partial x} = 0$, then, 2D LWR equation reduces to $\frac{\partial}{\partial y}\left(\rho v_y - D\frac{\partial \rho}{\partial y}\right) = 0$, solving this equation with symmetric boundary conditions $\rho(y = b) = \rho(y = -b)$ and inserting the centre condition $\rho(y = 0) = \rho_0$, the following steady-state equations are derived:

$$\rho(y) = \frac{\rho_{max}}{1+\left(\frac{\rho_{max}}{\rho_0}-1\right)e^{\left(\frac{2v_{yb}s_y e^{-\frac{b}{s_y}}}{D}\left(\cosh\left(\frac{y}{s_y}\right)-1\right)\right)}} \tag{18}$$

We generated and tested four different central conditions for steady-state lateral distribution such as $\rho(0) = 0.01\rho_{max}$, $\rho(0) = 0.5\rho_{max}$, $\rho(0) = 0.99\rho_{max}$ and $\rho(0) = \rho_{max}$. Figure 3 shows the analytical solution for steady-state, and numerical solutions for different time period and above-mentioned four cases.

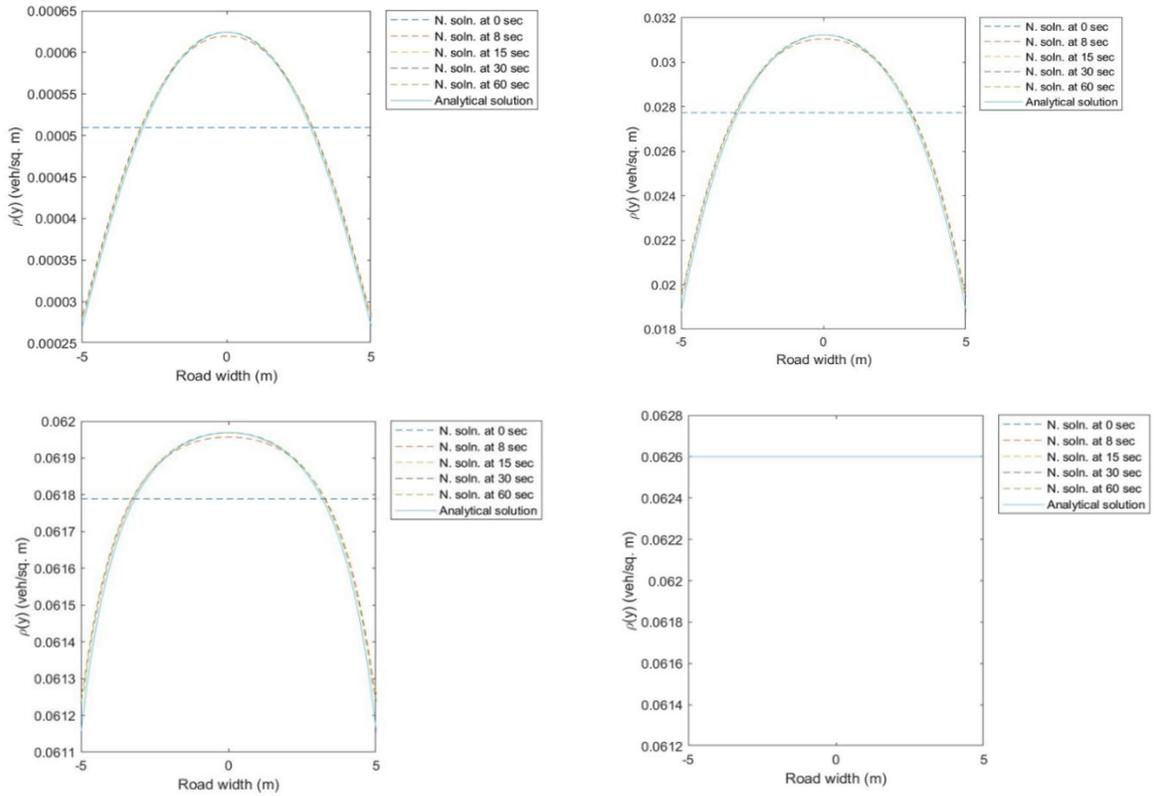

Figure 3: Analytical solutions for steady-state lateral density distribution and corresponding numerical solutions for the central conditions: $\rho(0) = 0.01\rho_{max}$, $\rho(0) = 0.5\rho_{max}$, $\rho(0) = 0.99\rho_{max}$ and $\rho(0) = \rho_{max}$

In all the cases, numerical solution converges to analytical solution and does not change over time and also, when the maximum density is reached (case 4), no vehicle can move away from the boundary as there is no space. These results show the model consistency, implementation and correctness of numerical scheme of lateral convection and diffusion terms.

Moreover, three cases are considered for lateral dynamics which has extreme values for density distribution profiles to check the model's robustness and plausibility. We argue that if the model behaves well in these extreme cases, it is expected to produce good results in general.

**Case 1: Lateral concave density distribution**

The initial condition of this scenario is that the traffic density is distributed in concave form with maximum density at the road centre. It is represented in the mathematical formation as follows:

$$\rho[x,y,0] = \rho_{max}\left(1 - \left(\frac{y}{b}\right)^2\right) \tag{19}$$

Here, it is assumed that the longitudinal dynamics does not vary, so only the lateral dynamics is relevant. The evolution of traffic density in two space dimensions over time is given in the first column of the plots of Figure 4. At time t=0 s, maximum density is at the road centre and about half the changes are performed within the first 3-5 s corresponding to typical lane-changing times, the traffic is dispersed laterally based on availability of lateral space, boundary repulsion speed and diffusion. After time is 8 s, there is no much variation in lateral dynamics (refer 5$^{th}$ figure in column 1), these results are consistent with lane changing dynamics of vehicles, for e.g., average lane changing duration vary from 4 s to 17 s (Toledo and Zohar 2007). The boundary force generated by $v_y$ prevails eventually leading to less densities at the boundaries.

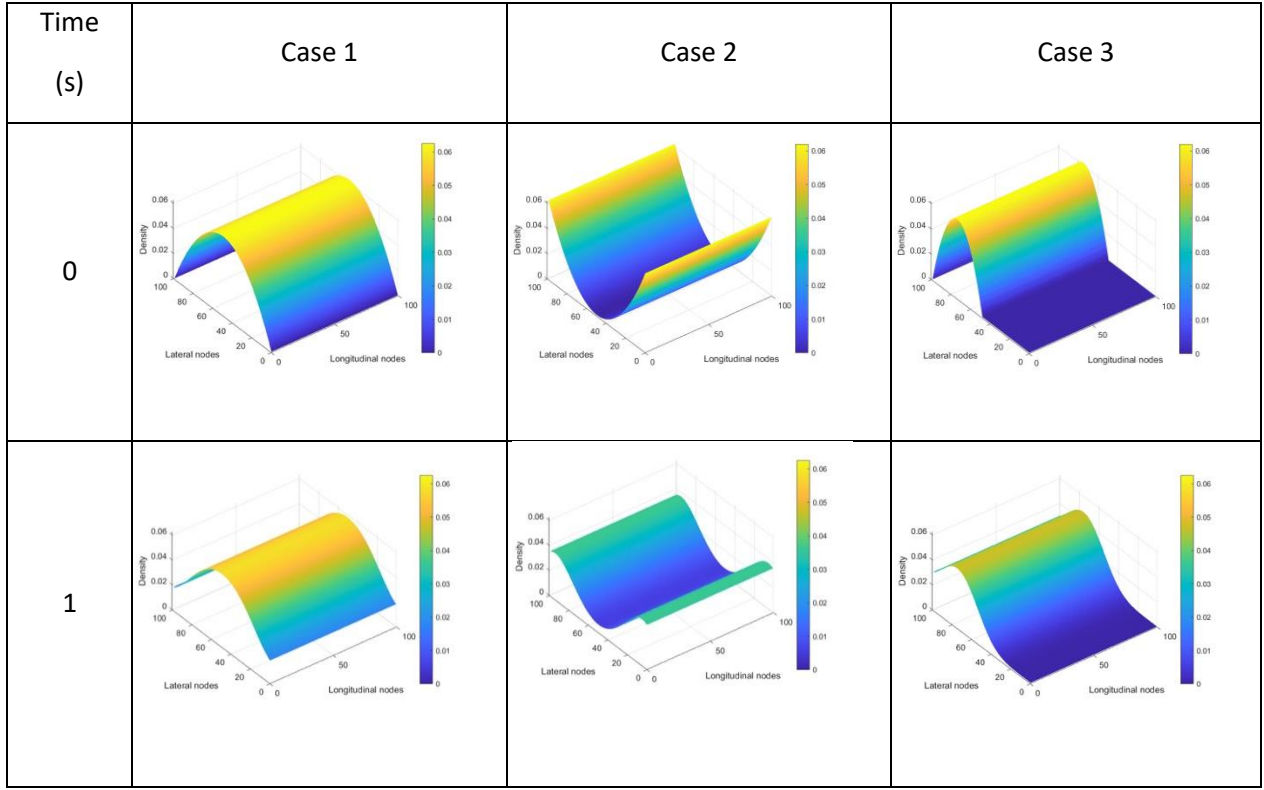

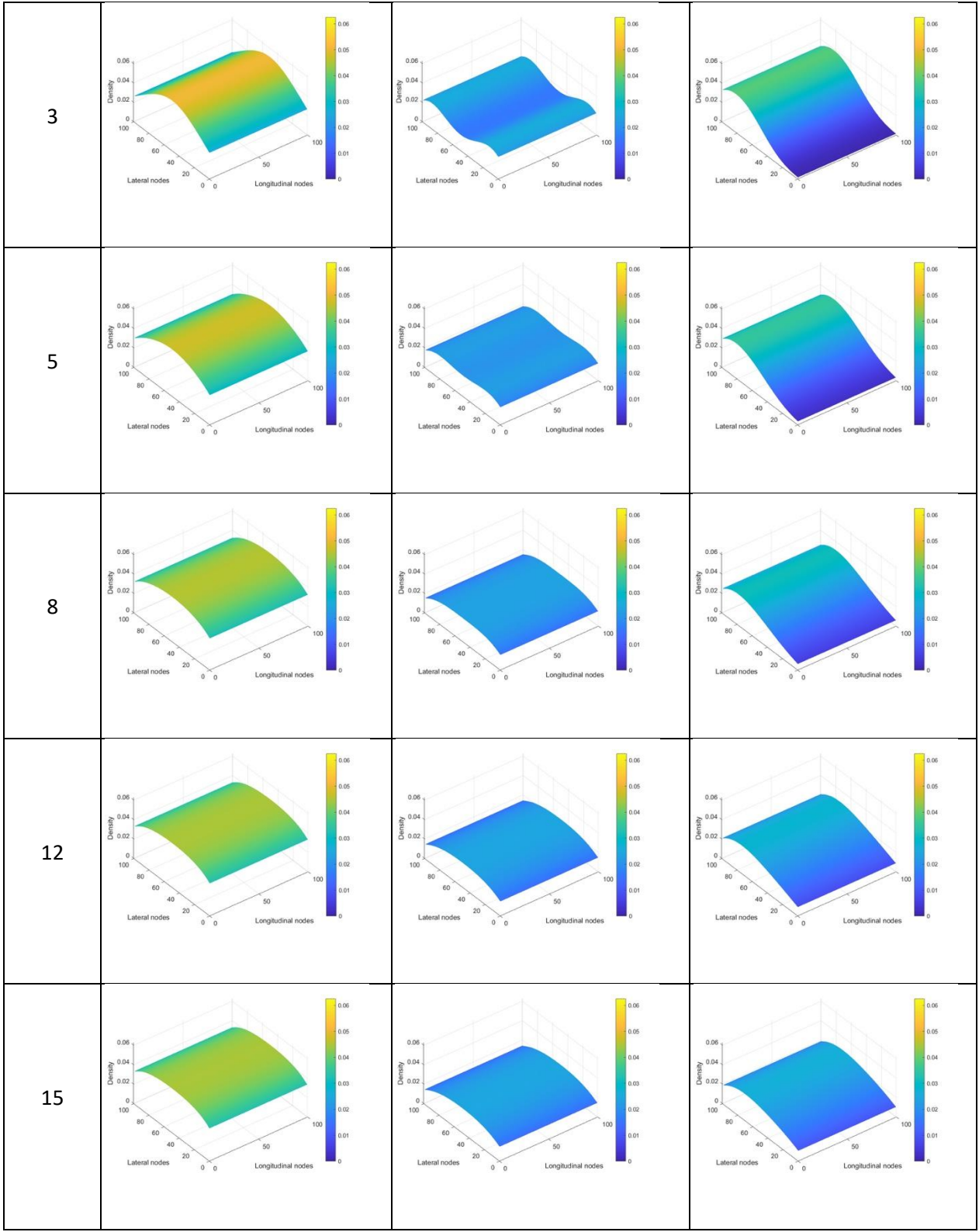

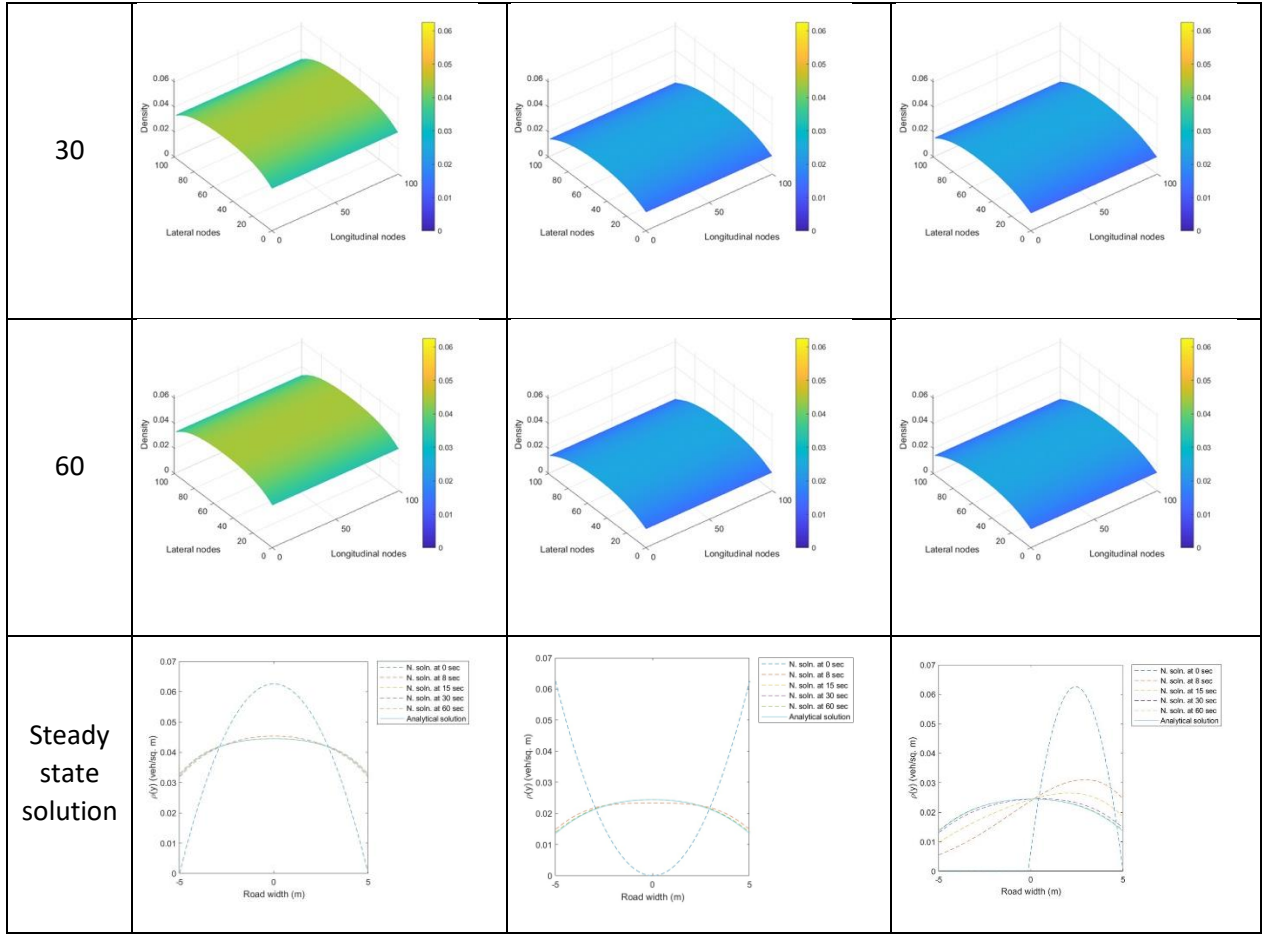

Figure 4: Evolution of traffic in two dimensions spatially over time - Lateral density distributions for three cases

**Case 2: Lateral convex density distribution**

In this case, density distribution is convex laterally in which initial conditions are minimum density at the road centre and maximum density at the road boundary which is represented as follows:

$$\rho[x, y, 0] = \rho_{max} \left(\frac{y}{b}\right)^2 \qquad (20)$$

Figures in column 2 of Figure 4 shows density distribution for two-dimensional planes at different time periods. Initially, at time t =0 s, density is maximum at the road boundary and at time t = 1-3 s, density is dispersed, and its distribution looks like a wave. The reason is that boundary has given negative repulsive force and hence, the vehicle tries to move away from the road boundary as soon as possible. At the same time, vehicles do not have enough time to reach the centre of the road from the boundary. Hence, the wave-like density profile was created. At time t = 8 s, lateral dispersion is almost stabilized (similar like lateral concave as in case 1) and after that there were no significant changes observed in the lateral density profile.

**Case 3: Lateral concave density distribution for half of the road**

In this scenario, the initial condition is such that traffic density is zero in half of the road section while the other half has concave density distribution with a maximum density at the centre of that section. It is mathematically represented as follows:

$$\rho[x, y, 0] = \begin{cases} \rho_{max}\left(1 - \frac{\left(y - \frac{b}{2}\right)^2}{\left(\frac{b}{2}\right)^2}\right), & if\ y \geq 0 \\ 0, & Otherwise \end{cases} \quad (21)$$

Figures in column 3 of Figure 4 represents the initial condition. It can be observed that from figure 2 of column 3, the vehicles are dispersing to the other side of the empty road section at time t = 1 s. At time t= 15 s, half of the vehicles move to the other side of the empty road (refer figure 7 of column 3). Near 30 s, density profile stabilizes afterwards there are no significant changes observed. Moreover, in all the cases, numerical solution converges to analytical solution for steady-state lateral density distribution (refer row 10 in Figure 4).

**Case 4: Shock waves and Rarefaction waves**

In the above-mentioned three cases, we have analysed lateral dynamics with diffusion terms. In this case, longitudinal dynamics is included to simulate the shock wave condition. The transition from lower density in upstream to higher density in downstream may result in the sudden jump of density causing the formation of a shock wave. To simulate such conditions, the following initial conditions are assumed:

$$\rho(x, y, 0) = \begin{cases} \frac{\rho_{max}}{5}, & if\ x \leq \frac{L}{2} \\ \rho_{max}, & Otherwise \end{cases} \quad (22)$$

At time t = 0 s, a shock wave occurs at x =L/2 and the corresponding initial density configuration in a 2D plane is shown in figures of row 1 in Figure 5. At t= 1 s, the shock wave is traveling upstream in the Greenshields' and Del Castillo and Benitez's models since shock wave speed is negative (refer Figure 1). At the same time in the lateral direction, vehicles accumulate and form a parabola like structure near the shock wave interface in the Greenshields' and Del Castillo and Benitez's models. This may be due to road boundary repulsion and jam density conditions in the downstream side (longitudinal direction) and other factors. It can be observed that at the second time, vehicles accumulation is emerging in the lateral direction at time t = 12 s in the Del Castillo and Benitez's model and at t = 30 s in the Greenshields' model. This may be because in the congested condition, the vehicle speed obtained by the Del Castillo and Benitez's model is lower compared to the Greenshields' model and hence, vehicles dissipate less in the jam region and create vehicles accumulation early.

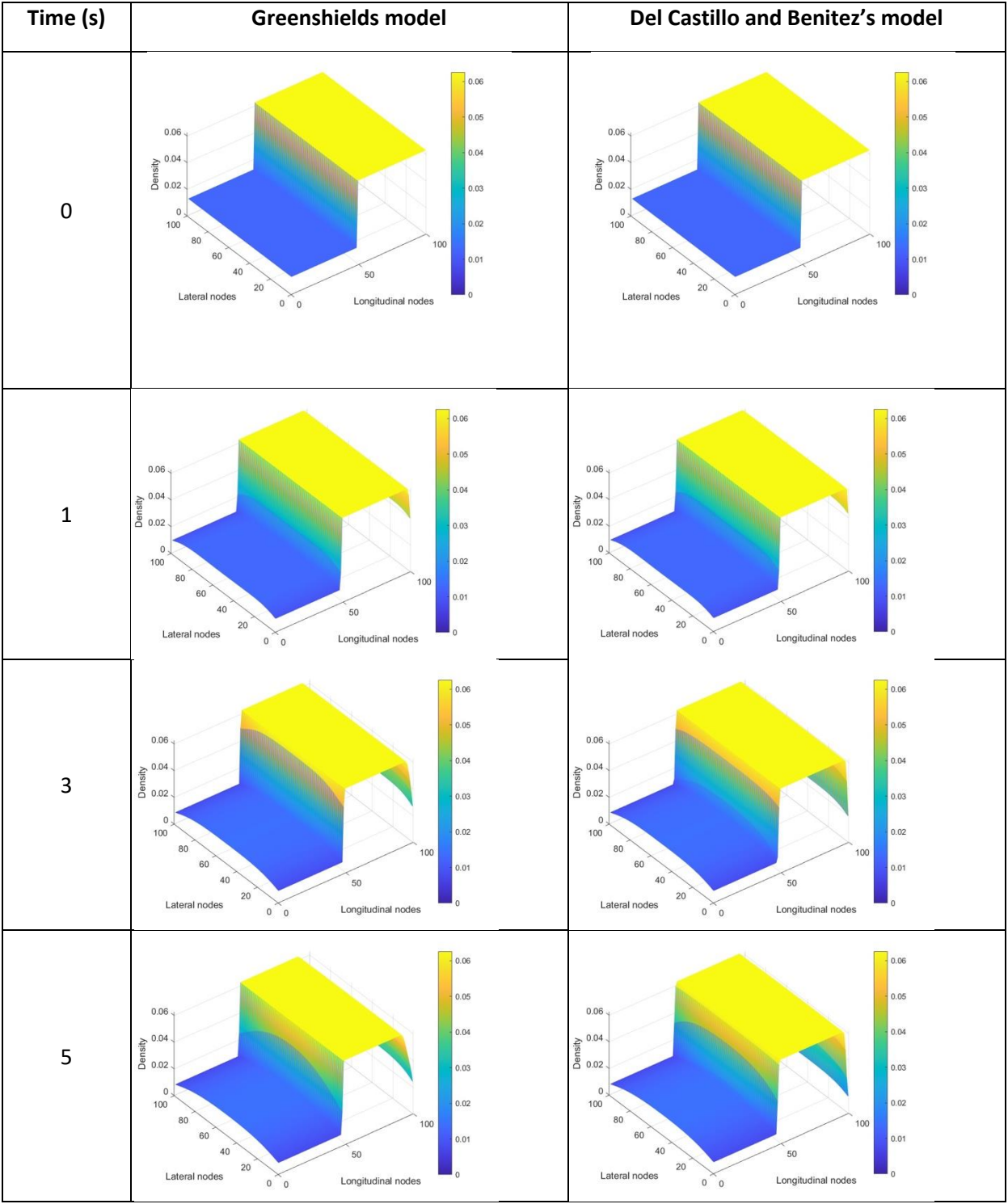

| 8 | | |
|---|---|---|
| 12 | | |
| 15 | | |
| 30 | | |

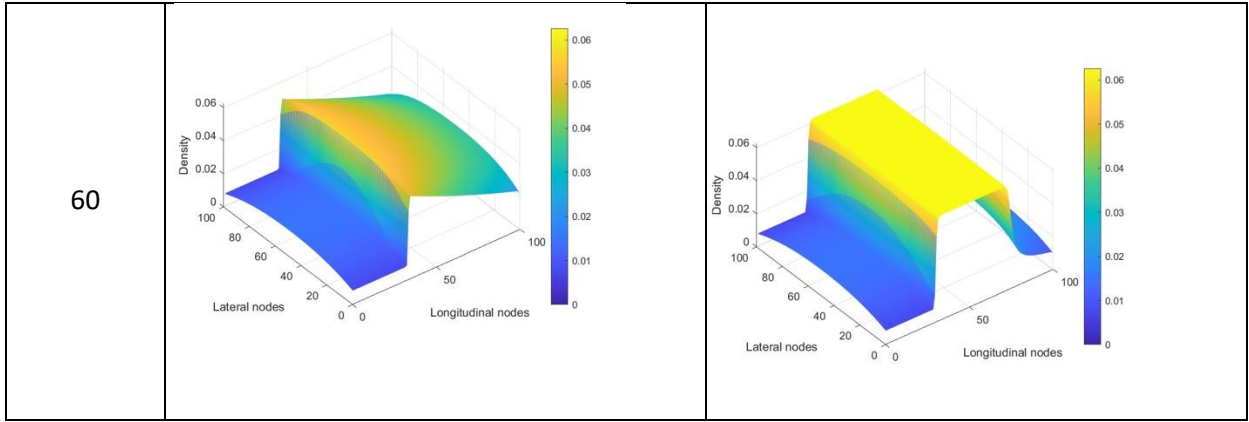

Figure 5: Evolution of traffic density in two dimensions spatially over time with shock wave and rarefaction wave

At time t = 3 s, rarefaction wave is clearly visible on the downstream side of the road section, and after that, it travels backward and jam density in the upstream direction dilutes at subsequent time intervals. Lateral direction vehicles disperse according to the dispersion coefficient and distance from the road boundary. Moreover, it is also numerically convincing since, when starting with a section of maximum density ("Rarefaction Wave"), of course, no lateral displacements are possible, i.e., the lateral boundary repulsion cannot take effect. Therefore, the yellow surface just remains a flat yellow surface until it vanishes under the dynamics of its' interfaces. At time t = 60 sec, in Greenshields' model, jam condition almost dilutes, and in Del Castillo and Benitez's model, it does not dilute fully. This is due to the low speed of vehicles in the congested condition, and the maximum flow-density is less (we assume that vehicles leave from the jam at maximum flow-density) in the Del Castillo and Benitez's model.

**Case 5: Bike Simulations**

In the aforementioned cases, extreme values for density distribution profiles in the lateral and longitudinal directions are considered assuming that vehicles prefer to travel in the center of the road. In this case, to evaluate lateral density dynamics of 2D LWR model, we considered a nontrivial example by simulating single-lane bike traffic considering the observed lateral density profile given by Guo et al. (2021). Figure 14 (Guo et al. 2021) shows the evolution of density in the polar radial direction in the ring-shaped track where the density is uniformly distributed initially and later the distribution converges to a localized profile. This is due to the asymmetry of the circular path and people tend to take the inner radius. This is simulated by introducing a constant speed bias $v_b$ towards the inner part of the bike path (right side) and modifying lateral speed, which is given by:

$$v_y = \left( v_{yb} exp\left(-\left(\frac{b+y}{s_y}\right)\right) - exp\left(-\left(\frac{b-y}{s_y}\right)\right) + v_{R_{bias}} \right)\left(1 - \frac{\rho}{\rho_{max}}\right) \qquad (23)$$

Table 3 shows the model parameters to simulate bicycle traffic. The evolution of density in the lateral direction over time is depicted in Figure 6. It was observed that the cyclist moved progressively towards the inner part of the track ($y = 1.5\ m$) over a period of time. The bicycles accumulate densely between the road's center and inner part of the track, and this density profile matches with experimental data conducted at Anqing Normal University China and this shows the real-world applications of the model.

Table 3: Model parameters for the biased case

| Parameter | Value |
|---|---|
| Road width, $2|b|$ | $3\ m$ |
| Lateral step-size, $\Delta y$ | $0.1\ m$ |
| Mesh, $N^x \times N^y$ | $1 \times 30\ cells$ |
| Time step-size, $\Delta t$ | $0.001\ s$ |
| Lateral boundary interaction, $s_y$ | $2\ m$ |
| Maximum boundary repulsion velocity, $v_{yb}$ | $0.2\ m/s$ |
| Lateral velocity bias, $v_{bias}$ | $0.1\ m/s$ |
| Coefficient of diffusion, $D$ | $0.01\ m^2/s$ |

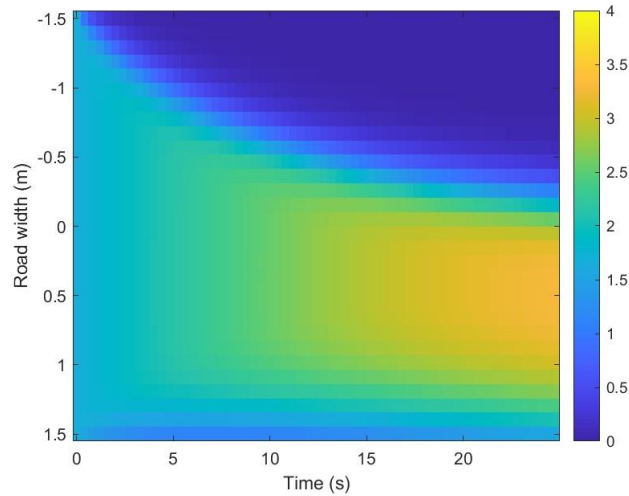

Figure 6: Evolution of bicycle lateral density distribution across the road cross-section

All the above cases show that the proposed model shows satisfactory results and is consistent with traffic dynamics. Different kinds of dispersion in the lateral direction can be achieved by adjusting the parameters such as road boundary repulsion, desired speed or diffusion coefficient solely and a combination of these.

## 5. Conclusions

In this study, we have proposed a first-order macroscopic 2D traffic flow model of the LWR type in order to represent both longitudinal and lateral macroscopic dynamics for disordered traffic flow. For this purpose, we have extended the continuity equation to two spatial dimensions reflecting conservation of vehicles in both directions. In the longitudinal direction, the continuity equation is closed by the Greenshields' and Castillo and Benitez's speed-density relation while a novel model of exponential boundary repulsion is applied to close the lateral dimension. In order to reflect the driver's desire to go to regions with a lesser density, a lateral Fick's law is adopted that leads to a lateral diffusion term that also helps suppress unwanted sharp transitions and lateral shocks. The final 2D first-order model consists of five parts: continuity equation, two hydrodynamic relations for the longitudinal and lateral flow-density components, and longitudinal and lateral speed equations.

The numerical scheme proposed to solve this 2D-LWR model is based on directional and operator splitting. The implementation and correctness of numerical scheme of lateral convection and diffusion terms is checked by comparing with analytical solution of lateral-steady state distributions for four different central conditions. The model was tested for different initial traffic flow scenarios representing a wide range of flow conditions such as concave and convex densities in the lateral direction, longitudinal shock and rarefaction waves. In all cases, the proposed model shows plausible results. Moreover, using the model, we simulated evolution of lateral density profile for a single-lane bicycle traffic, and the model able to reproduce asymmetry behavior in the inner lane path. However, a comprehensive investigation with data is necessary to calibrate and validate the model and to fully understand how disordered lane-free traffic flow evolves over space and time. As an extension of this model, we are currently developing second-order 2D macroscopic model for disordered traffic and these bring additional complexity to the model especially two speed equations and the lateral advective terms. Moreover, we are planning to extend the proposed model to a multi-class model to capture different vehicle types. It is also conceivable to adapt the model to other instances of directed two-dimensional flow such as pedestrian evacuation scenarios or planning Marathons or other mass-sports events which will be the subject of future studies.

## Acknowledgments

The first author is supported by the Prime Minister's Research Fellowship, Ministry of Education, Government of India at Indian Institute of Technology Kanpur.

32  M. Treiber, A. Kesting, Traffic Flow Dynamics: Data, Models and Simulation, Springer, Berlin (2013)

33  T. Toledo, D. Zohar, Modeling duration of lane changes, Transportation Research Record 1999 (1) (2007) 71-78.

34  E. F. Toro, Riemann Solvers and Numerical Methods for Fluid Dynamics: A Practical Introduction. Springer Science & Business Media. (2013).

35  D. Vikram, S. Mittal, P. Chakroborty, Stabilized finite element computations with a two-dimensional continuum model for disorderly traffic flow, Computers & Fluids. 232 (2022) 105205.

36  H. Wang, J. Li, Q. Chen, D. Ni, Representing the Fundamental Diagram: The Pursuit of Mathematical Elegance and Empirical Accuracy, Transport Research Board 89th annual meeting, Washington, DC, USA (2010).
24